# Hybrid integrated mode-locked laser diodes with a silicon nitride extended cavity

EWOUD VISSERS,[1,2,*] STIJN POELMAN,[1,2] CAMIEL OP DE BEECK,[1,2] KASPER VAN GASSE,[1,2] AND BART KUYKEN[1,2]

[1]*Photonics Research Group, Department of Information Technology, Ghent University IMEC, Ghent B-9000, Belgium*
[2]*Center for Nano- and Biophotonics (NB-Photonics), Ghent University, Ghent, Belgium*
**Ewoud.Vissers@UGent.be*

**Abstract:** Integrated semiconductor mode-locked lasers have shown promise in many applications and are readily fabricated using generic InP photonic integration platforms. However, the passive waveguides offered in such platforms have relatively high linear and nonlinear losses that limit the performance of these lasers. By extending such lasers with, for example, an external cavity the performance can be increased considerably. In this paper, we demonstrate for the first time that a high-performance mode-locked laser can be achieved with a butt-coupling integration technique using chip scale silicon nitride waveguides. A platform-independent SiN/SU8 coupler design is used to couple between the silicon nitride external cavity and the III/V active chip. Mode-locked lasers at 2.18 GHz and 15.5 GHz repetition rates are demonstrated with Lorentzian RF linewidths several orders of magnitude smaller than what has been demonstrated on monolithic InP platforms. The RF linewidth was 31 Hz for the 2.18 GHz laser.

## 1. Introduction

Mode-Locked Lasers (MLL) are comb lasers: The optical spectrum of these lasers contains multiple laser lines with a constant frequency spacing between them. Fiber based MLLs are already commercially available, and used in several applications such as astronomical spectrograph calibration [1], supercontinuum generation [2] or high speed low jitter electrical analog to digital converters [3]. Integrating such a MLL on a chip would allow to apply the lasers in a number of high volume applications such as distributed sensing of greenhouse gasses [4, 5], and optical ranging in autonomous vehicles [6]. Generic InP platform are very attractive for the integration as they offer all building blocks to construct mode-locked lasers on chip [7]. Indeed, monolithic chip based MLLs have already been demonstrated, but they have three drawbacks compared to fiber MLLs: The repetition rate is generally high (>10 GHz), and the Radio Frequency (RF) and optical linewidths are much higher than for fiber-based systems [4]. The high noise inhibits their use in a number of applications where the small form factor and the mass producibility would be very advantageous such as for coherent communication systems [8] and LIDAR [9]. Additionally, the high repetition rate is disadvantageous for (dual comb) spectroscopy applications as it negatively impacts the resolution [10].

All three of these disadvantages can be remedied simultaneously by increasing the cavity length of the MLL with a low loss cavity. This lowers the repetition rate of the MLL and simultaneously improves the RF and optical linewidth as the photon lifetime in the laser is increased [4]. The key necessity for linewidth improvements is that the extended cavity is low loss. Hence, current research efforts are trying to increase the cavity length without adding significant loss by using passive waveguides. This has already been demonstrated monolithically on a single chip using active/passive integration, where two different layer stacks are combined on the same InP chip, for separate active and passive waveguide sections on a monolithic InP platform [11]. However, the passive InP waveguides used for the external cavity in these monolithic lasers are moderately lossy (3 dB/cm), and they also suffer from two-photon absorption leading to sub-optimal performance.

To improve on the waveguide losses in the monolithic extended cavity MLLs, III/V-on-Si and III/V-on-SiN mode-locked lasers have also been demonstrated, where the different material systems are integrated using bonding or micro-transfer printing [12, 13]. In these types of lasers, the external waveguide cavity is made using a different material system with lower propagation losses than the passive waveguides in InP platforms, and the III/V gain medium is integrated by micro-transfer printing an amplifier on top of the waveguide circuit. A disadvantage of lasers using micro-transfer printing are the limitations on gain section design, and the moderate thermal performance of the active section, limiting the achievable output power in these lasers.

Hybrid integration by butt-coupling is a more flexible approach that can be used to couple external waveguide cavities to active gain sections. In this hybrid integration approach, two separate chips are fabricated that have waveguides that are extending to the edge of the chip. Those waveguides are then butt-coupled together. Using this approach, the gain sections can be made without any limitations imposed by a micro-transfer printing or bonding process e.g. in terms of active material and wavelength, and the feedback circuit can be made in any material system. Although the coupling section induces excess loss, this can be kept below 5 dB. Using the hybrid integration approach, a single mode laser has already been demonstrated with an intrinsic optical linewidth of only 40 Hz by using a SiN feedback circuit and an InP active chip [14].

A hybrid integration approach has also been used to decrease the footprint and power usage of integrated Kerr comb sources. In these devices, a III/V Reflective Semiconductor Optical Amplifier (RSOA) is coupled to an external cavity with a wavelength filter followed by a high Q ring resonator. In this way, systems with only electrical input ports and an optical comb output port have been demonstrated at 194 GHz [15] and 99 GHz [16]. However, scaling Kerr comb generation to lower repetition remains difficult, as the threshold power is inversely proportional to the repetition rate [17].

In this work, two mode-locked lasers are demonstrated, both using hybrid integration of an InP active chip and a SiN extended cavity chip. Their repetition rates are 15.8 GHz and 2.18 GHz.

## 2. Design and Fabrication

The laser structures described in this work are assembled using two different photonic integrated circuits. An InP RSOA chip made in the generic SMART Photonics platform is used for the active parts of the laser. The second chip consists of SiN waveguides fabricated with e-beam lithography and serves as the extended cavity for the laser. The laser has a self-colliding pulse configuration, where the saturable absorber is adjacent to the high reflectivity mirror [18].

The active InP chip consists of a multimode interference reflector (MIR) [19] at one end of the cavity. Immediately next to this is a saturable absorber section of 50 µm, followed by a gain section of 850 µm. This is followed by a 1x1 MMI acting as single mode filter, a 7° bend, and an angled output facet for butt-coupling to the SiN cavity chip. The latter minimizes reflections at the edge. The III/V chip schematic can be seen on the right in Fig. 1.

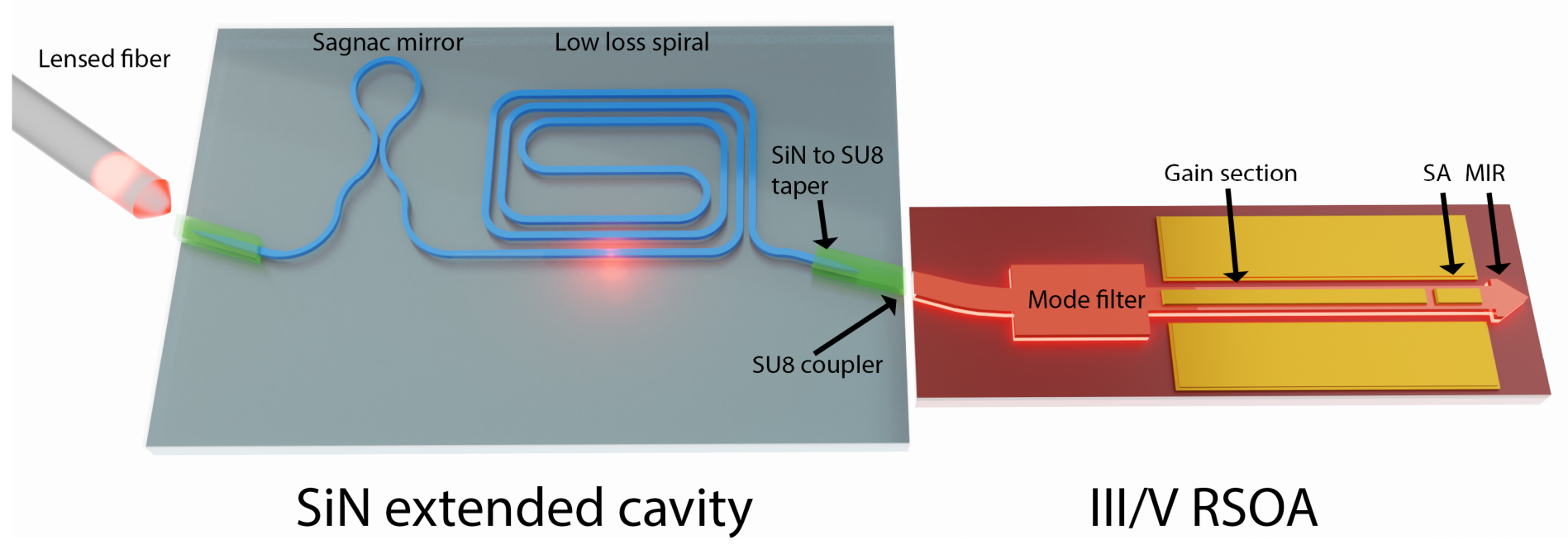

Fig. 1. Artistic impression of the laser. The outcoupling lensed fiber can be seen on the left, followed by the SiN extended cavity in the middle and the III/V RSOA on the right.

The chip with the SiN optical circuitry is fabricated by etching the 300 nm SiN LPCVD layer which is resting on top of a 3.3 µm $SiO_2$ layer on top of a 300 µm thick silicon handle wafer. The SiN waveguides are air-clad, and are defined using e-beam lithography and a subsequent RIE etch step. A tolerant and efficient edge coupling structure acting as a mode converter was made using the polymer SU8. A layer of SU8 was spin coated on the chip and waveguides were defined in the SU8 using contact lithography. Fig. 2 shows the overlap between the optical mode in the III/V waveguide and the SU8 waveguide at the edge of the chip for different chip alignment offsets at the fabricated dimensions, and for different waveguide dimensions at perfect alignment. During fabrication, an SU8 layer thickness of 875 nm was measured, resulting in a simulated mode overlap of 90 percent between both chips. The layer thickness is not very critical. As long as the layer is thicker than 0.8 um (mode cutoff), and thinner than 1.5 um, the overlap between both modes will be between 85 % and 90 % for the optimal width. At the fabricated size, the butt-coupling alignment tolerance has a 500 nm horizontal range and a 250 nm vertical tolerance to stay between 85 % and 90 % overlap between both modes.

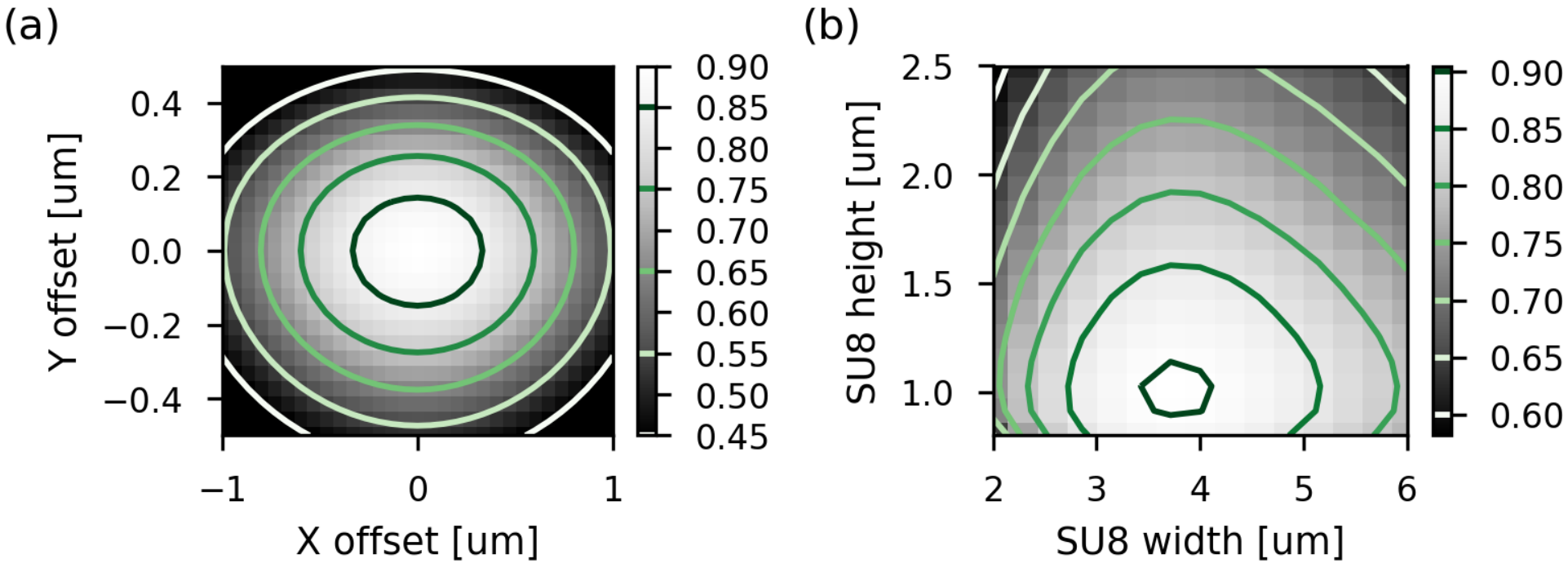

Fig. 2. Simulations showing the tolerances of the coupling between both chips. The simulated mode overlap between the III/V mode and the SU8 mode is shown as: (a) a function of distance from the optimal point at the fabricated thickness of 875 um and as: (b) a function of SU8 waveguide width and height.

The SU8 waveguide is 3.8 µm wide and has an angled facet a (15.2° to the normal) to match the emission angle of the InP chip and reduce reflection back into the waveguide. The fundamental TE mode of this waveguide has a 90% overlap to the fundamental output mode of the InP chip. The taper structure between the SiN- and SU8-waveguide can be seen in Fig. 3. Underneath the SU8, an inverse taper is made in the SiN to adiabatically transfer the optical mode from the SiN to the SU8 waveguide. The taper structure is followed by a SiN waveguide (1.2 mm for the high repetition rate laser, 3 cm for the low repetition rate laser), and a Sagnac reflector with a reflectivity of 70%. The Sagnac reflector was realized using a directional coupler made using SiN waveguides and a loop with a bend radius of 150 µm. The outcoupling is done using a second mode converter with an SU8 waveguide of 3 µm wide to horizontally match a lensed fiber mode as closely as possible. The height of this SU8 waveguide is the same as used for butt coupling to the InP chip, since it's made in the same processing step. The waveguide facets are made by cleaving the chip trough the SU8 waveguides. The SiN chip schematic can be seen in the middle of Fig. 1.

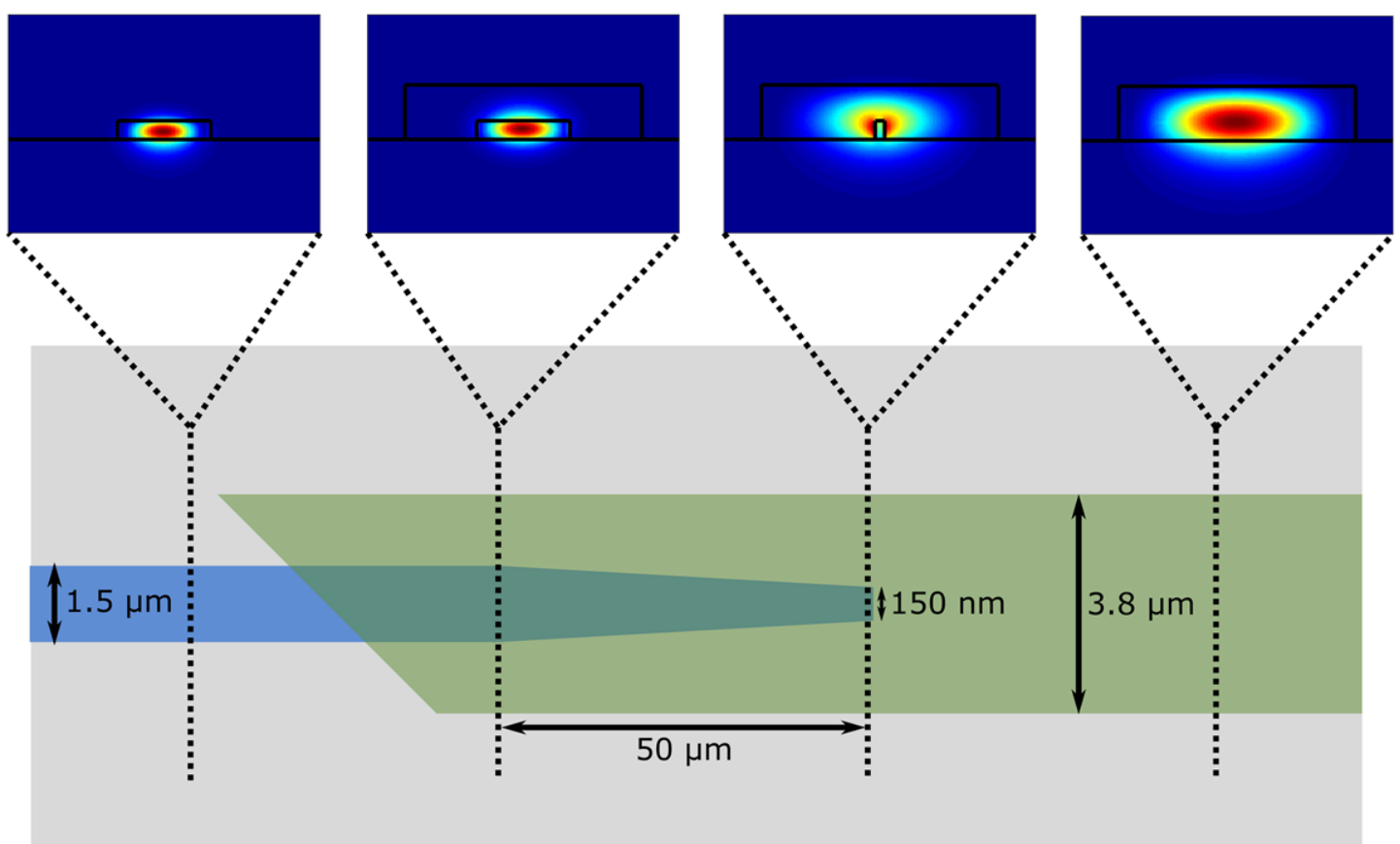

Fig. 3. Schematic of the taper structure, with the fundamental mode at certain positions. The blue SiN waveguide is on the left of the image, the green SU8 waveguide on the right.

The design of the SiN external cavity chip is very portable. Since the SU8 waveguide can be tuned in width and height, efficient coupling should be possible to a large variety of gain chips, regardless of the mode size. Furthermore, it can be used for operation in a wide wavelength range owing to the large transparency window of SU8 and SiN [20, 21]

## 3. Characterization

For the electrical contacting and temperature control of the InP chip, a printed circuit board (PCB) was made. Both the top and bottom of the PCB have an exposed copper pad, perforated with vias to increase the thermal conduction between both pads. A thermistor and the bottom of the InP chip were bonded to this copper pad using silver epoxy for a good thermal interface. The electrical connection between the InP chip and the PCB was made using wirebonds. A thermo-electric cooler (TEC) element was mounted between the bottom of the PCB and an aluminum heatsink, using thermal paste to ensure a good thermal interface. This heatsink was fixed to an optical table.

The SiN chip was placed on an XYZ arrangement of computer-controlled PI Q.545 piezo stages. The lensed fiber used for outcoupling was mounted on another set of the same piezo stages. Looking through a microscope, the two chips were roughly aligned. After that the alignment was further improved by optimizing for maximum current through the SA while applying a gain current and reverse bias to the gain section and SA respectively.

All measurements were done with the PCB's copper pad kept at 18 °C. The temperature was controlled using an Arroyo 6340 ComboSource driving the TEC element while using the thermistor for feedback. Two lasers with a different extended cavity length were characterized. Laser A has a 1.2 mm external cavity, and laser B has a 3 cm external cavity. Combined with the length of the RSOA and coupler sections, the repetition rates are 15.5 GHz for laser A and 2.18 GHz for laser B. The light from the laser was collected using a lensed fiber. Optical and RF spectra were taken at the same time. Using a 90:10 splitter, 10% of the light was sent to an Anritsu MS9740A OSA to measure optical spectra and power, and 90% was sent to a Discovery DSC10H photodiode. This photodiode was connected to an Agilent N9020A MXA. To measure autocorrelation traces, an EDFA was connected to the 90% arm to amplify the signal, followed by a polarization controller and an APE PulseCheck 150 autocorrelator (AC). For laser B, the setup was the same, except the 90:10 splitter was replaced with a 99:1 splitter. The measurement setup can be seen in Fig. 4 (a).

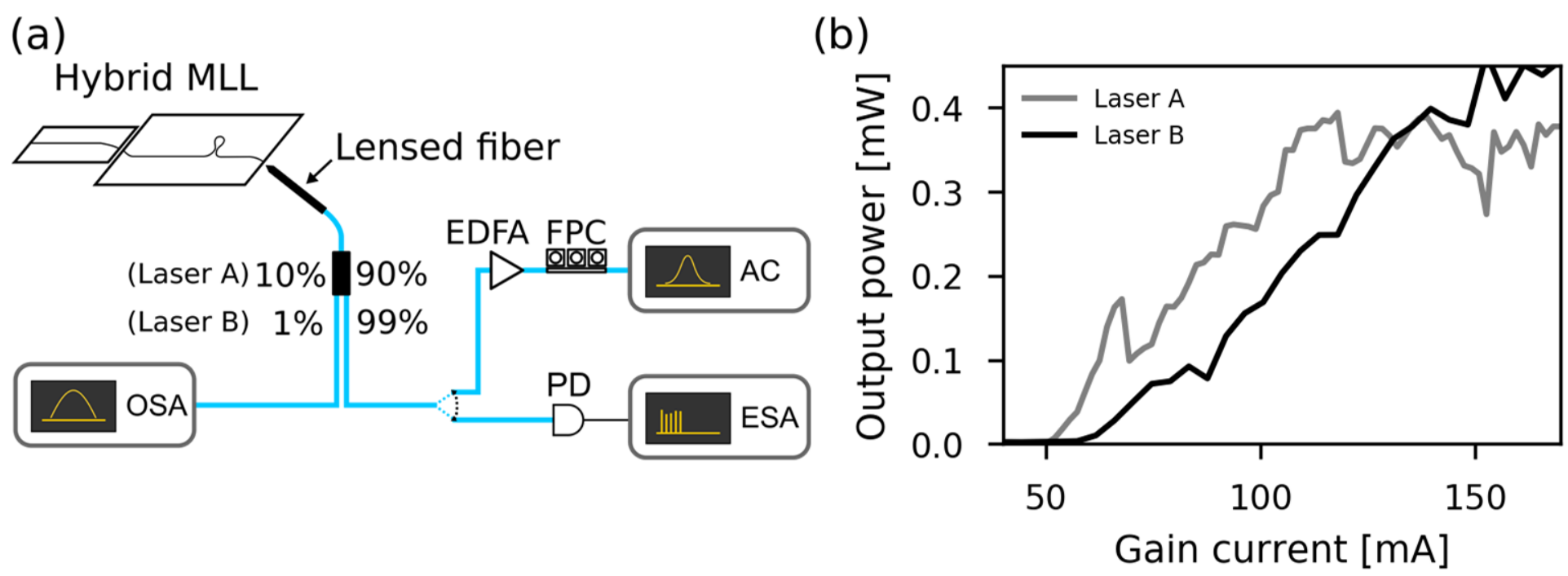

Fig. 4. (a) Measurement setup for laser A and B, with a different splitter used between both measurements. (b) LIV curves for both laser A and laser B for an absorber bias of 0 V.

The LI curves of both lasers are shown in Fig. 4 (b). It can be seen that laser B has a higher threshold current than laser A, which is expected due to the extra waveguide losses in the longer cavity. The saturation power of the lasers is around 0.4 mW optical output power in the lensed fiber. This is limited by the RSOA used in the laser, because the SMART Photonics platform is not optimized for high saturation powers, but rather for high gain SOAs. By replacing the RSOA with one optimized for high saturation powers, this saturation power can be increased.

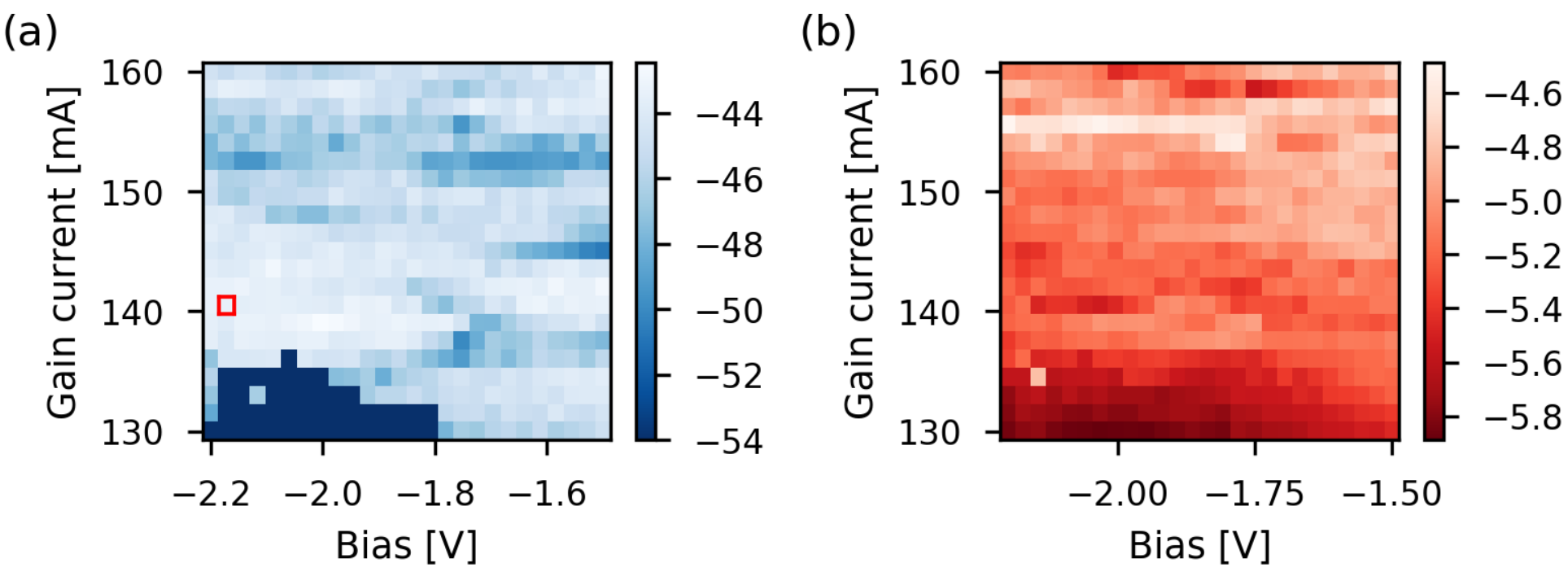

Fig. 5. Measurements for laser A. (a) shows the RF peak power and (b) the optical output power in the lensed fiber in dBm as a function of gain current and absorber bias. The red marker in (a) shows the optimal operating point with the narrowest RF linewidth.

Fig. 5 shows the RF peak power and optical output power measured for laser A at several absorber biases and gain currents. Almost the entire measured region shows a high RF power at 15.5 GHz. The narrowest RF linewidth was measured at a gain current of 140 mA and a SA bias of -2.16 V. The optical spectrum, the RF note measurement and an autocorrelation trace for these operating parameters are shown in Fig. 6. The optical spectrum shows a -3 dB bandwidth of 1.5 nm. A Voight profile was fit to the RF measurement, which indicates an estimated -10 dB linewidth of 940 KHz. This is half the RF linewidth of a similar laser made using active/passive integration in InP at a similar repetition rate of 16 GHz [22]. A $sech^2$ fit of the autocorrelation trace indicates a pulse length of 3.1 ps.

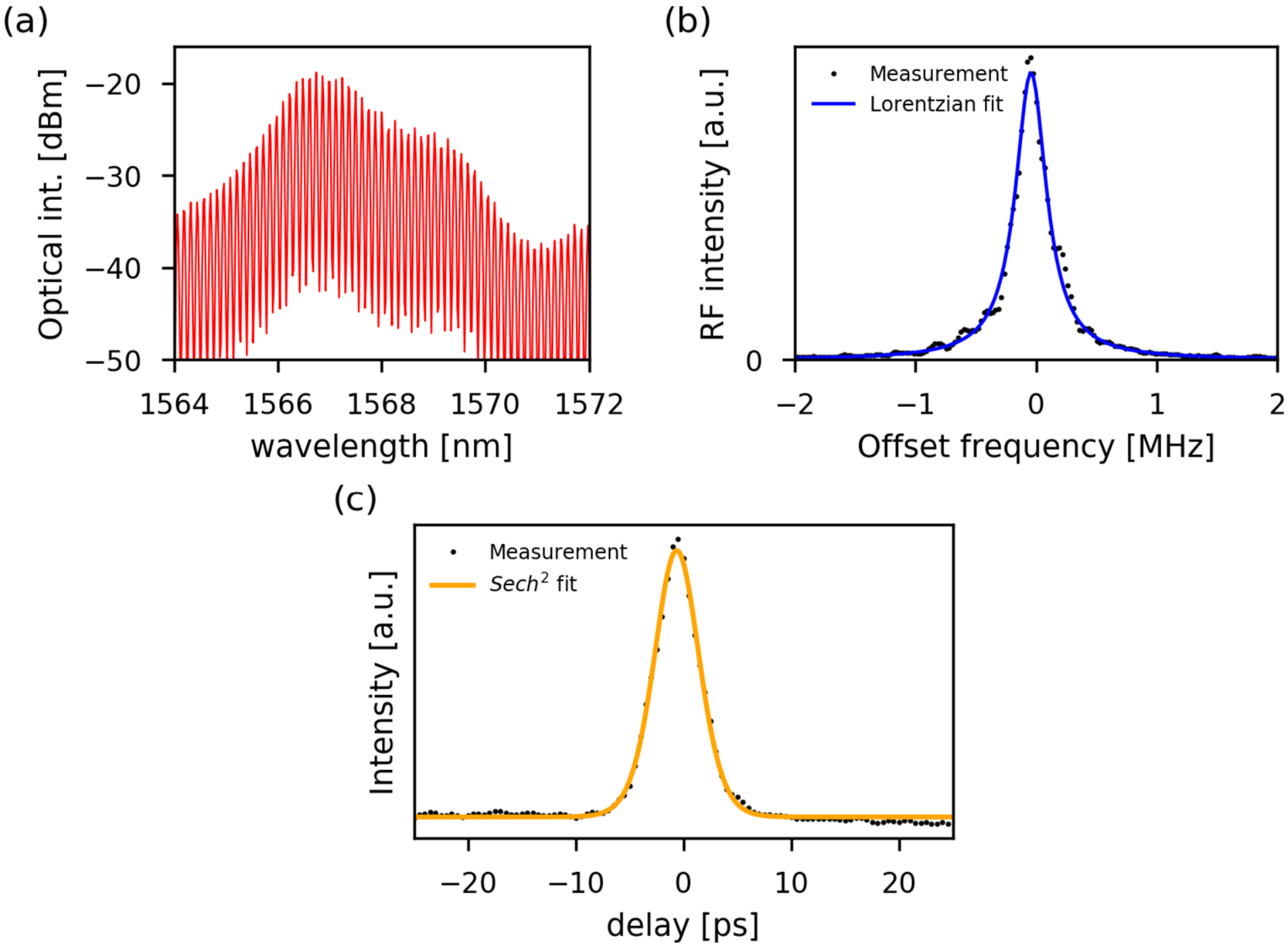

Fig. 6. Measurements at the optimal operating point for laser A. (a) shows the RF peak power and (b) the optical output power in the lensed fiber as a function of gain current and absorber bias. (c) shows the optical spectrum for laser A, (d) shows the RF linewidth measurement and the fitted Voight profile, (e) shows the autocorrelation and the fitted 3.1 ps $sech^2$ pulse.

Fig. 7 shows the RF peak power and frequency, and the optical output power for laser B at several absorber biases and gain currents. Since the repetition rate of laser B is 2.18 GHz, several harmonics can be measured within the bandwidth of the measurement equipment. From these harmonics, three pulsing operating regimes can be found: Q-switching, mode-locking at the fundamental frequency, and mode-locking at twice the fundamental frequency with residual mode-locking at the fundamental frequency.

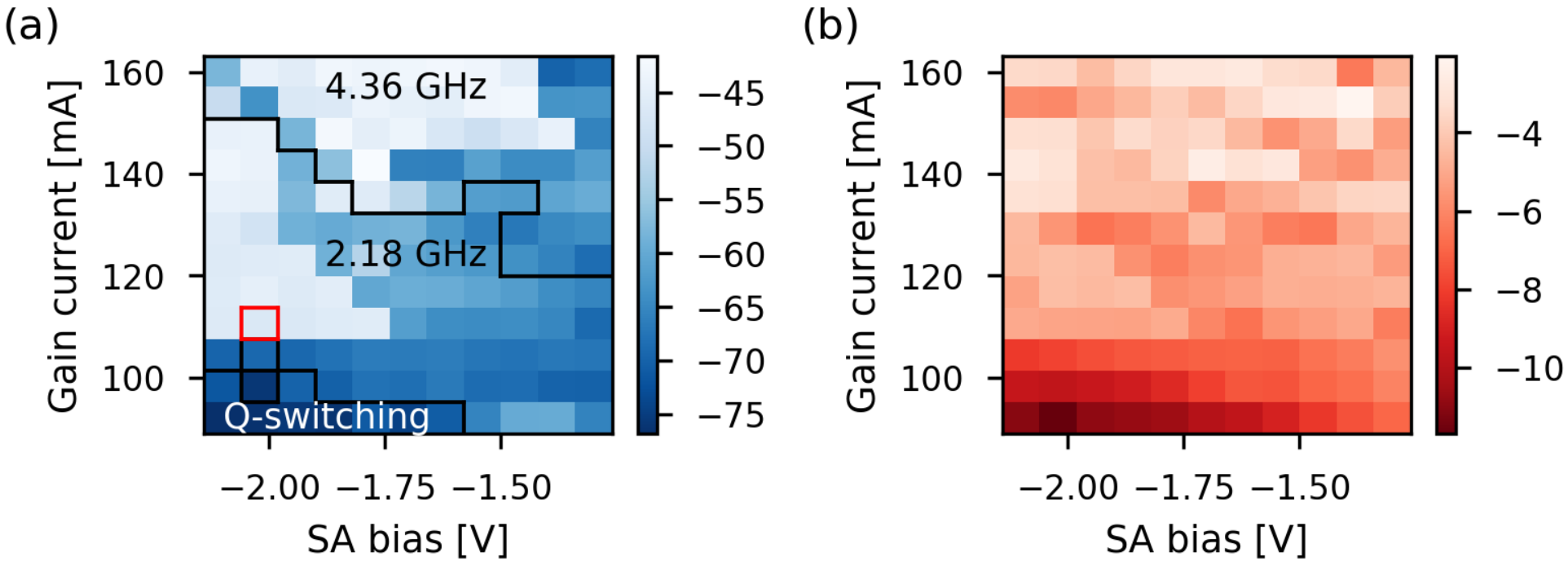

Fig. 7. (a) The RF peak height in dBm of the strongest note and (b) the optical output power in the lensed fiber in dBm as a function of gain current and absorber bias. The black lines in (a) separate the operating regions of Q-switching, and fundamental and second harmonic mode-locking. The red marker shows the optimal operating point with the narrowest RF linewidth.

The operating parameters leading to the narrowest RF linewidth were found to be a gain current of 111.1 mA and an SA bias of -2.04 V, in the region of fundamental frequency mode-locking. Since the RF linewidth at this point is lower than 1 kHz, a single sideband phase noise (SSB-PN) measurement was performed to measure the linewidth more accurately. This measurement can be seen in Fig. 8 (c). A Lorentzian fit gives a linewidth of 31 Hz. The optical spectrum at this point can be seen in Fig. 8 (a), with a -3 dB bandwidth of 4 nm. However, as can be seen from the spectrum, the comb is rather flat-top, and has a -10 dB bandwidth of 8.3 nm, containing 465 comb lines. The autocorrelation trace at this operating point can be seen in Fig. 8 (d). A $\mathrm{sech}^2$ fit leads to a pulse width of 6.31 ps. The combination of these measurements implies that the pulse is not transform limited. A pulse duration of 6.31 ps would lead to a -3 dB bandwidth of 0.4 nm using the $\mathrm{sech}^2$ time bandwidth product of 0.315. This is 10 times smaller than the measured bandwidth. Possibly, the pulse duration could be shortened by sending the pulse trough an (on-chip) dispersive element.

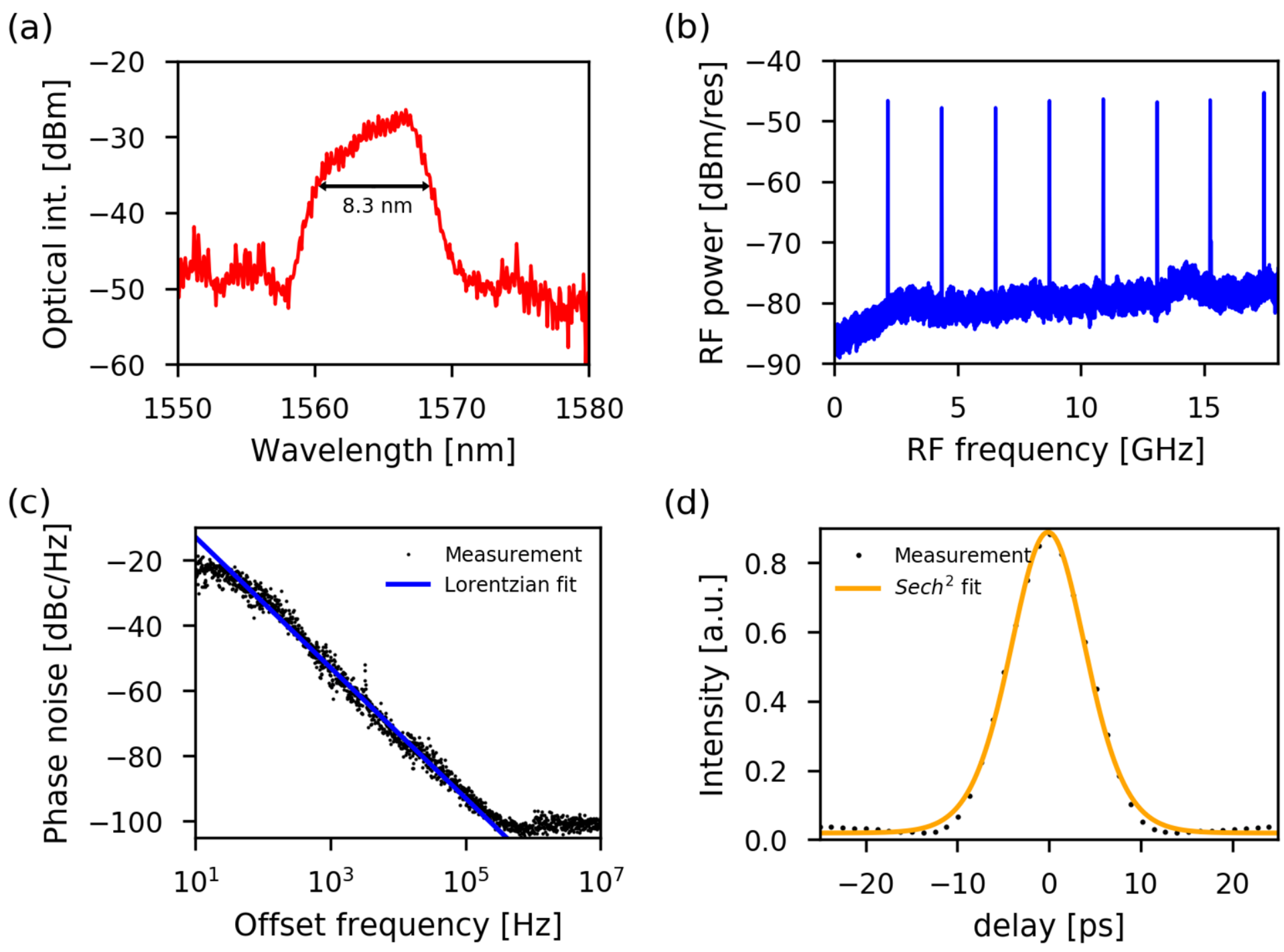

Fig. 8. (a): Measurements at the optimal operating point for laser B. (a) shows the optical spectrum, (b) the RF spectrum. (c) shows the single sideband phase noise measurement of the fundamental RF peak and the fitted Lorentzian profile with a width of 31 Hz. (d) shows the autocorrelation trace and the fitted 6.31 ps $sech^2$ pulse.

## 4. Conclusion

We have demonstrated for the first time two butt-coupled integrated mode-locked lasers using an active InP chip made in the SMART Photonics platform and an external cavity chip made in SiN. One laser had a 15.5 GHz repetition rate, the other a 2.18 GHz repetition rate. The 2.18 GHz laser has a low RF linewidth of 31 Hz, which is several orders of magnitude below linewidths achieved using the active/passive integration approach. The coupling scheme used is theoretically very portable, so it can be used to fabricate (mode-locked) lasers using other RSOAs as well.

Furthermore, both lasers were assembled using the same RSOA, where only the passive cavity was swapped. Therefore, the change in performance from laser A to laser B can solely be attributed to the passive cavity. This proves that the RF linewidth of an on-chip mode-locked laser can be reduced by only changing the external cavity for a longer one when using a low-loss waveguide platform such as SiN.

Finally, using this integration strategy, a large variety of integrated mode locked lasers across many different material platforms and wavelengths can be fabricated. This approach to integrate mode-locked lasers opens the door to a wide range of applications such as dual-comb-OCT and spectroscopy [10, 23]

**Funding.** H2020 Marie Skłodowska-Curie Actions (MICROCOMB - 812818), European Research Council (ELECTRIC – 759483), Fonds Wetenschappelijk Onderzoek (1S54418N and 12ZB520N)

**Acknowledgements.** We acknowledge SMART Photonics for the production of the active III/V chips, and LioniX International for the SiN wafer material.

**Disclosures.** The authors declare no conflicts of interest.

**Data availability.** No data were generated or analyzed in the presented research.